\documentclass[runningheads]{llncs}
\usepackage[T1]{fontenc}
\usepackage{graphicx}
\usepackage{booktabs}
\usepackage{wrapfig}
\usepackage[cmex10]{amsmath}
\usepackage{subfigure}
\usepackage{subcaption}   
\usepackage{amssymb}
\usepackage{algorithmic}
\usepackage{array}
\usepackage{mdwmath}
\usepackage{mdwtab}
\usepackage{eqparbox}
\usepackage{url}
\usepackage{hyperref}
\usepackage{algorithm}
\usepackage{algorithmic}
\usepackage{epstopdf,cite,color}

\newcommand{\beq}{\begin{equation}}
\newcommand{\eeq}{\end{equation}}
\newcommand{\beqn}{\begin{eqnarray}}
\newcommand{\eeqn}{\end{eqnarray}}
\newcommand{\beqno}{\begin{eqnarray*}}
\newcommand{\eeqno}{\end{eqnarray*}}
\newcommand{\bma}{\begin{displaymath}}
\newcommand{\ema}{\end{displaymath}}
\newcommand{\bnu}{\begin{enumerate}}
\newcommand{\enu}{\end{enumerate}}
\newcommand{\bce}{\begin{center}}
\newcommand{\ece}{\end{center}}
\newcommand{\btb}{\begin{tabular}}
\newcommand{\etb}{\end{tabular}}

\begin{document}
\title{SpoofTrackBench: Interpretable AI for Spoof-Aware UAV Tracking and Benchmarking}
%
%
\author{Van~Le\inst{1}\orcidID{0009-0007-5638-7178} \and
Tan~Le\inst{2}\orcidID{0000-0003-0807-4357} }
\authorrunning{V. Le et al.}
\institute{Virginia Polytechnic Institute and State University, Blacksburg, VA 24061, USA. \\
\email{vanl@vt.edu} \and
Hampton University, Hampton, VA 23669, USA\\ 
\email{tan.le@hamptonu.edu}\\
\url{https://sites.google.com/site/thanhtantp}
}

\maketitle

\begin{abstract}
SpoofTrackBench is a reproducible, modular benchmark for evaluating adversarial robustness in real-time localization and tracking systems (RTLS) under radar spoofing. Leveraging the Hampton University Skyler Radar Sensor dataset, we simulate drift, ghost and mirror-type spoofing attacks and evaluate tracker performance using both Joint Probabilistic Data Association (JPDA) and Global Nearest Neighbor (GNN) architectures. Our framework separates clean and spoofed detection streams, visualizes spoof-induced trajectory divergence, and quantifies assignment errors via direct drift-from-truth metrics. Clustering overlays, injection-aware timelines, and scenario-adaptive visualizations enable interpretability across spoof types and configurations. Evaluation figures and logs are auto-exported for reproducible comparison. SpoofTrackBench sets a new standard for open, ethical benchmarking of spoof-aware tracking pipelines, enabling rigorous cross-architecture analysis and community validation.


\end{abstract}

\section{Introduction}

Radar-based tracking systems play a vital role in surveillance, autonomous navigation, and aerial asset monitoring. As these systems become increasingly integrated into civilian and defense infrastructure, they face rising threats from adversarial spoofing. Techniques such as drift, ghost and mirror injection can distort detection streams, mislead assignment logic, and destabilize tracker performance—posing safety, security, and operational risks~\cite{Kerns2014, UAVSurvey2025,le2025HDQNN}. In high-stakes environments like UAV coordination or border surveillance, even minor assignment errors can lead to mission failure or critical misinformation~\cite{Sathaye2022}.

Despite growing awareness of spoofing risks, there is a lack of standardized, reproducible benchmarks that rigorously evaluate radar trackers under adversarial conditions. Most existing trackers are tested in clean environments with synthetic clutter or noise, but without deliberate, labeled spoof injections. Comparisons across architectures—such as classical probabilistic trackers like Joint Probabilistic Data Association (JPDA) and graph-based trackers like modified Global Nearest Neighbor (GNN)—remain fragmented and unreproducible due to inconsistent scenario setups, injection logic, and evaluation metrics.

While progress has been made in spoof detection for domains such as GNSS~\cite{Psiaki2016,Rados2024} and Wi-Fi~\cite{Chen2009}, radar-based tracking/RTLS benchmarks remain modality-specific and lack unified testing protocols. Existing work rarely incorporates visual interpretability (such as cluster overlays, trajectory continuity plots, or spoof annotation tools), which are crucial for understanding failure modes and reasoning about adversarial robustness. Moreover, most evaluations focus solely on tracker accuracy without considering reproducibility under randomized seeds, separation of clean and spoofed logs, or modularity for rapid scenario switching.

We present \textbf{SpoofTrackBench}, a modular, reproducible, and visually interpretable benchmark framework for radar-based tracking/RTLS in adversarial scenarios. SpoofTrackBench simulates radar detection streams with controlled injection of spoof types, offering labeled spoof logs, toggleable 2D/3D visualization modes, and tracker-agnostic evaluation harnesses. It supports both JPDA and GNN architectures, allowing bidirectional comparison under identical spoof conditions. The framework is built around robust logging, randomized scenario reinitialization, and interpretability overlays—making it suitable for both academic benchmarking and system-level diagnostic analysis.

Our proposed framework offers the following core contributions:
1) \textbf{Modular Spoof Injection Logic:} Parameterized drift, ghost and mirror spoofing, supporting rapid injection cycling and labeled detection streams.
2) \textbf{Quantitative Drift and Assignment Metrics:} Calculation of trajectory divergence, assignment misalignment, and spoof-induced cluster disruption for both JPDA and GNN trackers.
3) \textbf{Cross-Tracker Comparison Harness:} Unified metric logging and scenario reinitialization enabling fair, reproducible comparison.
\vspace{-0.1in}
\subsection{Related Work and Literature Background}
\vspace{-0.1in}
\textbf{Radar Spoofing and Adversarial Tracking}:  
Growing interest in adversarial resilience has spurred research into spoof detection across radar, GNSS, and wireless modalities. Das et al.~\cite{das2019gnss} examined Kalman-based GNSS trackers under spoofed satellite signals, revealing assignment drift but lacking control over synthetic scenario variability. Gupta et al.~\cite{gupta2024wifi} proposed interpretability techniques for Wi-Fi-based spoof detection, though their system did not track spatial cluster continuity or quantify assignment divergence in radar domains. Li and Mohapatra~\cite{li2020spoof} advanced spoof classification metrics for single-modality pipelines but omitted cross-architecture evaluations or trajectory distortion metrics.  
These efforts highlight the fragmentation in current literature: most studies focus on spoof detection, not on spoof-aware tracking performance, and rarely support reproducible comparison across tracker types or spoof injection schemas.

\noindent\textbf{Benchmarks in Related Domains}:  
Benchmarks such as KITTI \cite{Geiger2012KITTI} and MOTChallenge \cite{Dendorfer2021MOTChallenge} have shaped evaluation practices in computer vision and autonomous tracking. However, these datasets assume clean detection environments or focus on occlusion, leaving adversarial spoofing unaddressed. Few benchmarks simulate labeled injection scenarios, and fewer still include control over spoof parameters, separation of clean/spoofed logs, or interpretability overlays critical for understanding tracker failure modes. Radar-specific benchmarks remain scarce, especially those that support multiple spoof types, scenario modularity, and quantitative drift-from-truth metrics.

\noindent\textbf{Need for Reproducible and Interpretable Adversarial Benchmarks}:
Across these domains, a recurring limitation is the absence of benchmarks that provide:
1) Modular spoof injection for diverse attack types (drift, ghost, mirror),
2) Tracker-agnostic evaluation with standard logging interfaces, 
3) Visual overlays for interpretability and cluster inspection,
3) Reproducible scenario cycling and randomized seed management.
SpoofTrackBench addresses these gaps through a unified framework designed for scientific rigor, extensibility, and practical clarity.
\vspace{-0.1in}
\subsection{Contributions}
\vspace{-0.1in}
\begin{wrapfigure}{r}{0.45\textwidth}
\vspace{-0.6in}
    \centering
    \includegraphics[width=0.35\textwidth]{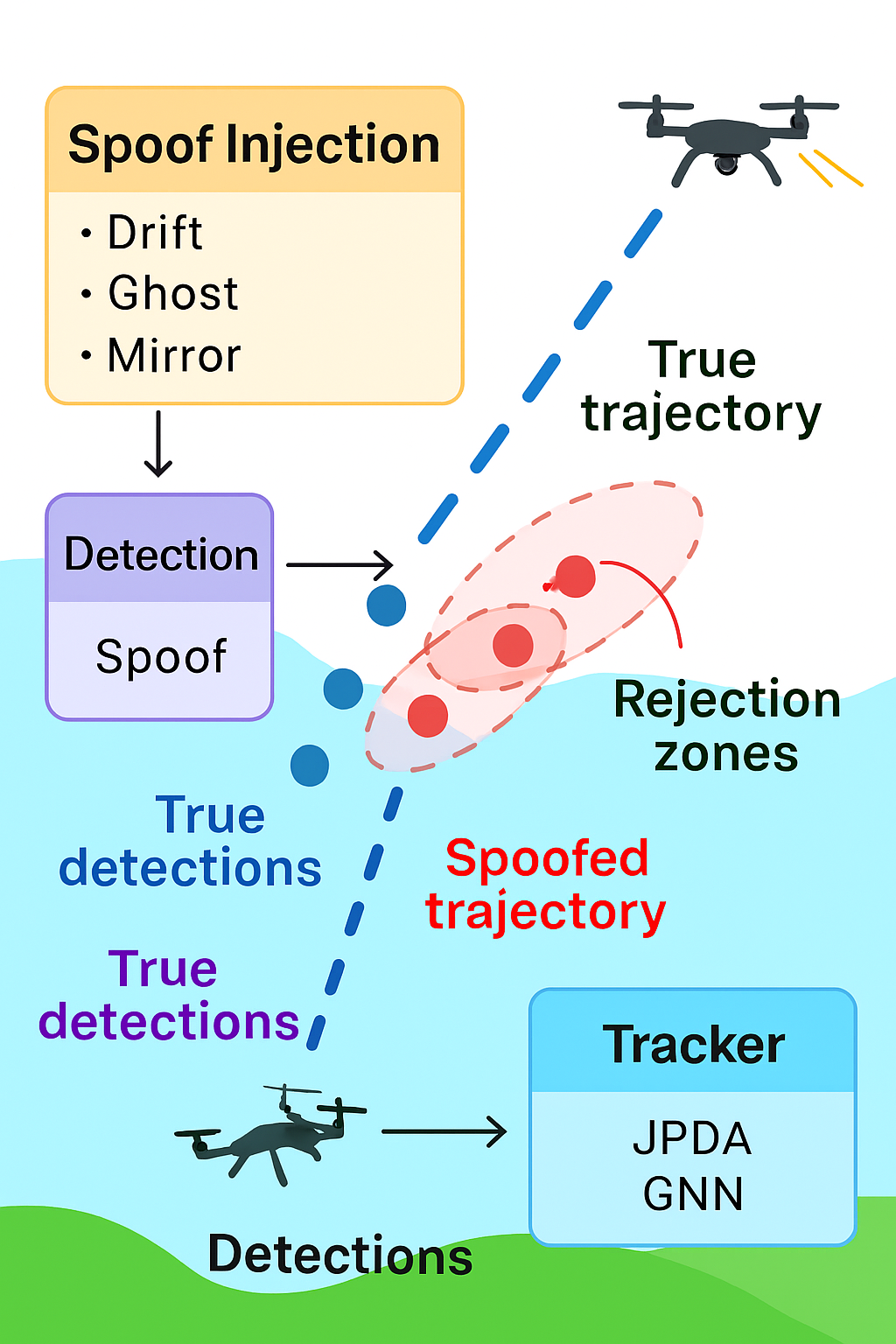}
\vspace{-0.1in}
\caption{Schematic comparison of spoof-aware trajectory tracking. JPDA and GNN respond differently to adversarial spoofing within gating regions: JPDA dilutes spoofed detections probabilistically, while GNN applies threshold-based rejection. The illustration highlights trajectory drift, spoof infiltration, and gating logic under adversarial conditions.}
    \label{schematic}
    \vspace{-0.3in}
\end{wrapfigure}
We present SpoofTrackBench in Fig. \ref{schematic}, a reproducible benchmarking framework for radar-based tracking under adversarial spoofing conditions. It enables scientific comparison of tracker architectures, spoof injection types, and interpretability routines through modular simulation and standardized metrics. Specifically, our contributions include:\\
\noindent 1) Modular Spoof Injection Logic: 
A configurable system supporting drift, ghost, and mirror spoof types with scenario-adaptive controls. Spoofs are injected using reproducible functions, enabling batch simulation and consistent labeling for interpretability and diagnostics.\\
\noindent 2) Reproducible Drift Quantification and Evaluation: 
Separation of clean and spoofed detection streams, randomized seed cycling, and structured logging across time-steps. We quantify drift-from-truth, assignment divergence, and cluster misalignment under controlled adversarial scenarios.\\
\noindent 3) Tracker-Agnostic Benchmark Harness:
A unified evaluation pipeline comparing JPDA and GNN trackers under identical spoof conditions.\\
\noindent 4) Dataset Integration and Extensible Design:
SpoofTrackBench supports modular scenario definitions and logging for reusability. The framework is designed for extensibility into future spoof types, multi-sensor fusion, and intelligent tracker architectures.
\section{System Model}
\label{System}
\vspace{-0.1in}
We consider a multi-sensor aerial surveillance architecture designed for spoof-aware tracking in contested environments \cite{lefevre2014survey, bar2001tracking}. The system comprises distributed sensing modalities feeding into a centralized AI-based Fusion Center, which performs adversarially resilient assignment, trajectory reconstruction, and interpretability logging.
\vspace{-0.1in}
\subsection{Sensor Network and Data Ingestion}
\vspace{-0.1in}
Sensor modalities include multiple sensors and/or detectors.
1) \textbf{Skyler Radar Sensors}: Provide high-resolution detection of non-cooperative aerial targets through micro-Doppler signatures. Particularly sensitive to platform dynamics and spoof-induced anomalies.
2) \textbf{ADSB (Automatic Dependent Surveillance–Broadcast)}: Capture positional broadcasts from cooperative aircraft, enabling spoof contrast through ID-grounded trajectories.
3) \textbf{LTE/4G Signal Detectors}: Record RF emissions and packet timings from airborne platforms, supporting detection continuity in signal-rich environments and complementary spoof resolution.
Each sensor independently logs detections at fixed intervals. These detection logs are streamed to a central fusion node, where sensor-origin tags are preserved, malformed entries are filtered, and missing timesteps are padded using modular error-handling routines.
\vspace{-0.1in}
\subsection{Complementarity and Robustness through Fusion}
\vspace{-0.1in}
Each sensor contributes distinct signatures. Firstly, radar enables platform dynamics and spoof traceability.
Next, ADSB anchors cooperative identity and continuity.
Finally, LTE/4G signals offer temporal density and spoof correlation through RF behavior.
Fusion across these channels yields greater resilience to spoof injection. Disagreement between sensors (e.g. radar-based spoof detection without ADSB corroboration) triggers interpretability overlays and trajectory warnings. The centralized fusion logic exploits sensor complementarity to maximize assignment integrity and spoof-awareness.
\section{Benchmark Design}
\label{BenchmarkDesign}

SpoofTrackBench demonstrated in Fig. \ref{schematic} is designed to support reproducible, spoof-aware benchmarking of radar tracking architectures under modular and adversarial conditions. The benchmark is structured into distinct components spanning data ingestion, spoof injection, tracker evaluation, and interpretability-driven logging \cite{xiang2015learning}.
\vspace{-0.1in}
\subsection{Radar Dataset and Scenario Setup}
\label{DatasetSetup}
\vspace{-0.1in}
We utilize the Hampton University Skyler Radar Sensor dataset, which provides volumetric micro-Doppler returns from airborne targets under varied maneuvering conditions. Detection frames are sampled across timesteps at configurable update rates (e.g. 0.5–1.0s), allowing realistic simulation of both cooperative and non-cooperative aerial behavior.

Simulation scenarios are modularized as MATLAB structures, enabling parameter toggling for:
\begin{itemize}
    \item Number and type of platforms (e.g. UAV, fixed-wing)
    \item Sensor coverage and detection uncertainty
    \item Scenario geometry, spoof entry points, and time durations
\end{itemize}

Each scenario instantiates synthetic ground truth with associated clean detection logs, which are subject to downstream spoof injection and tracking.
\vspace{-0.1in}
\subsection{Spoof Injection Logic}
\label{SpoofInjection}
\vspace{-0.1in}
SpoofTrackBench supports three primary spoof types, i.e. \textbf{Drift Spoof}, \textbf{Ghost Spoof} and \textbf{Mirror Spoof}. 
The first type of \textbf{Drift Spoof} is the result of temporal deviation of detection from true location, which simulates the gradual trajectory corruption.
Next, \textbf{Ghost Spoof} is the case of false detections with no physical platform origin, which challengs the tracker resilience and assignment purity.
Finally, \textbf{Mirror Spoof} is the reflection-based symmetry across geographic axes or sensor grid, which is  designed to mislead motion models.
We can present the derivations for these three spoof types as follows:
\begin{itemize}
    \item \textbf{Drift Spoof}:
    Let the true detection at time $t$ be $\mathbf{x}_t = [x_t, y_t]^\top$. Drift spoofing introduces a temporal deviation such that the spoofed detection becomes $\tilde{\mathbf{x}}_t = \mathbf{x}_t + \Delta_t$, where $\Delta_t = \alpha t \cdot \hat{\mathbf{v}}$, $\hat{\mathbf{v}} = \text{unit drift direction}$.
    This simulates gradual trajectory corruption over time.
    \item \textbf{Ghost Spoof}:
    Ghost spoofing injects detections $\tilde{\mathbf{x}}_t$ at time $t$ with no corresponding platform or track origin: $\tilde{\mathbf{x}}_t \notin \{ \mathbf{x}_t^{(i)} \mid i = 1, \dots, N \}$, and $\text{label}(\tilde{\mathbf{x}}_t) = \text{spoof}$.
    These detections challenge tracker assignment logic and clutter rejection.
    \item \textbf{Mirror Spoof}:
    Given a true detection \( \mathbf{x}_t = [x_t, y_t]^\top \), mirror spoofing reflects it across a defined axis (e.g. \( x = x_0 \)) or grid symmetry:
$\tilde{\mathbf{x}}_t = [2x_0 - x_t, y_t]^\top$.
    This misleads motion models by simulating symmetric but false trajectories.
\end{itemize}
Scenario parameters dictate spoof type, injection rate, and platform targets. 
The function reinitializes detection logs per run, applying randomized seeds for controlled variability. Clean and spoofed detection streams are stored separately with spoof flags and timestamps for traceability.
\vspace{-0.1in}
\subsection{Tracker Architectures}
\label{Trackers}
\begin{figure}[ht!]
\vspace{-0.3in}
    \centering
    \includegraphics[width=0.5\textwidth]{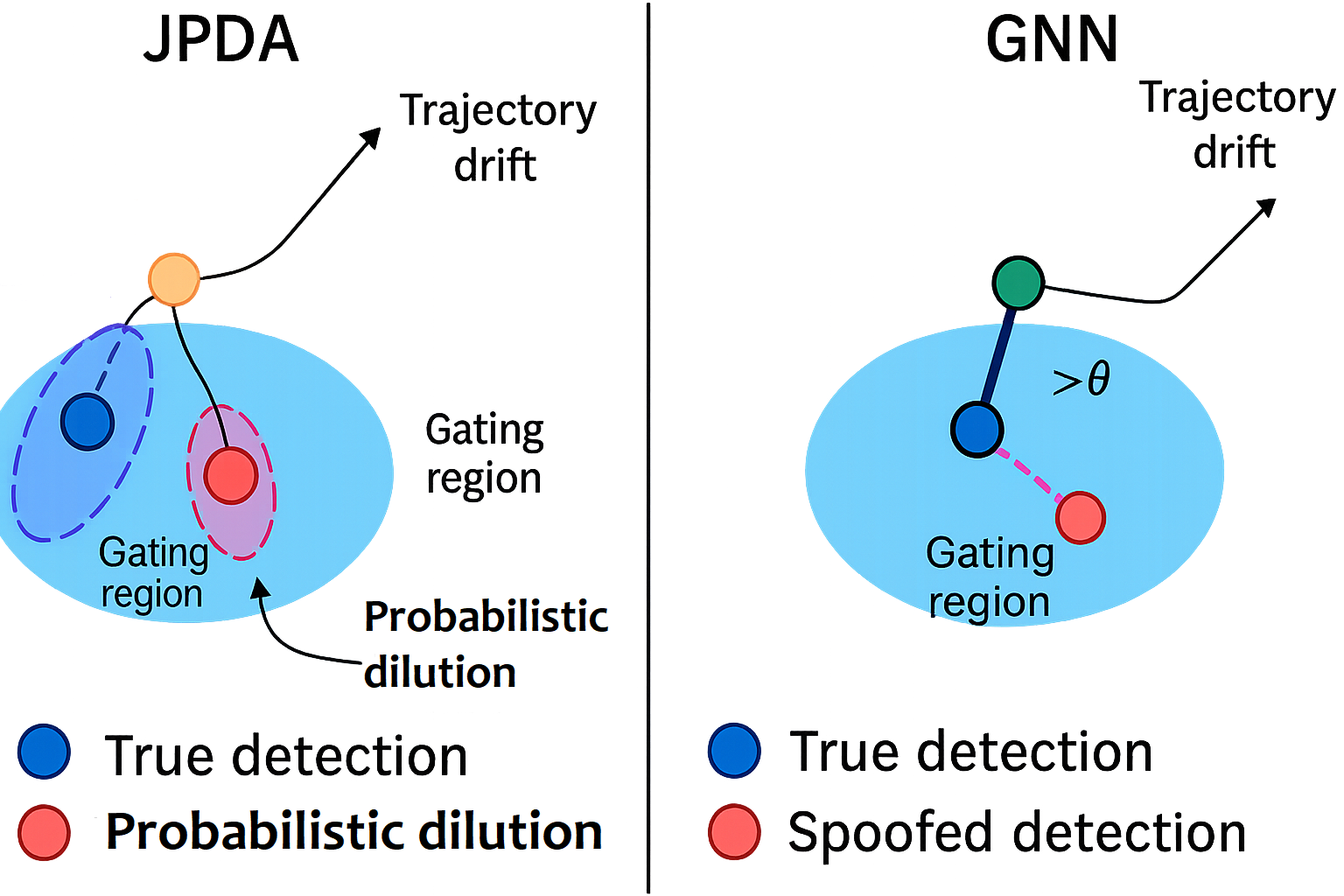}
    \vspace{-0.1in}
\caption{Comparative schematic of JPDA and GNN tracking under spoofing conditions. JPDA handles spoofed detections via probabilistic dilution within the gating region, while GNN applies threshold-based rejection to isolate spoofed inputs. The illustration highlights differences in trajectory drift response and spoof filtering mechanisms.}    \label{Comparative_figure}
\vspace{-0.3in}
    \end{figure}

\vspace{-0.1in}
SpoofTrackBench supports tracker evaluation across JPDA and GNN (see Fig. \ref{Comparative_figure}). 
Here, JPDA supports gating thresholds, detection clustering and assignment scoring, whilst detections in GNN are nodes and edges encode spatial-temporal affinity.
\noindent \textbf{Joint Probabilistic Data Association (JPDA) Tracker}:
The JPDA tracker performs soft data association by computing the probability that each detection originates from a given track. At each time step \( t \), for track \( j \), the update is a weighted sum over all gated detections:
\beqn
\hat{\mathbf{x}}_t^{(j)} = \sum_{i=1}^{M} \beta_{ij} \cdot \mathbf{z}_t^{(i)}
\eeqn
where $\mathbf{z}_t^{(i)}$ is the $i$-th detection at time $t$, $\beta_{ij} = P(\mathbf{z}_t^{(i)} \mid \text{track } j)$ is the association probability, $M$ is the number of detections within the gating region, and $\hat{\mathbf{x}}_t^{(j)}$ is the updated state estimate for track $j$.
JPDA accounts for missed detections and clutter by normalizing over all feasible association events. It is sensitive to spoofing when false detections fall within the gate, as they dilute the association probabilities and degrade track confidence.

\noindent \textbf{Global Nearest Neighbor (GNN) Tracker}:
The GNN tracker performs deterministic data association by assigning each detection to the nearest track based on a cost metric, typically Mahalanobis distance. At each time step $t$, the tracker solves:
\beqn
\min_{\sigma \in \mathcal{S}} \sum_{i=1}^{N} d(\mathbf{z}_t^{(i)}, \mathbf{x}_t^{(\sigma(i))})
\eeqn
where $\mathbf{z}_t^{(i)}$ is the $i$-th detection at time $t$, $\mathbf{x}_t^{(j)}$ is the predicted state of track $j$, $d(\cdot, \cdot)$ is the association cost (e.g. Mahalanobis distance), $\sigma$ is a permutation mapping detections to tracks, and $\mathcal{S}$ is the set of feasible assignments.
GNN uses gating to reject unlikely associations and solves the assignment problem using algorithms like the Hungarian method. It is robust to clutter and spoofing when spoofed detections fall outside gating thresholds.

Both architectures share input formats and evaluation pipelines, ensuring consistent comparison. 
We need to configure the parameters of gating distance or gate threshold, clutter density and birth probability $P_{\mbox{birth}}$, and tracker confidence thresholds via scenario definitions, which is then passed into tracker initialization modules. 
This enables robust tuning across spoof conditions.
\vspace{-0.1in}
\subsection{Enhancement with AI-based Target Tracking}
\vspace{-0.1in}
We derive the simulation for object tracking by using the radar sensor in the radar surveillance system. Then, we develop the algorithms of multiple target tracking using the global nearest neighbor. In the algorithm, we utilize the clustering the region and address the process of data association. It means that we will fit the single most likely hypothesis at each scan. There are many challenges in the cluster-based multi-target tracking. The consecutive errors would come from the cause of the wrong measurements for the specific target or the false alarm. This usually happens when we consider the scenario of tracking multiple objects. Also, the clustering methods may create the wrong region of object movement and hence cause the incorrect acquired data. All of these lead to false association between the previous known targets and measurements. 
Hence, the system cannot keep track on the target (i.e. loosing track or breaking track) due to the wrong assignment of measurements to track. 
Specifically, the cluster must be designed correctly so that we must avoid the case of high density clusters. In this case, the computation for tracking cause a heavy burden to the computing system and may lead to serial numbers of false tracks.
Furthermore, we need to design the fair clusters with the same/similar number of targets to be tracked. 
By doing so, we can balance the computing resource allocation to every cluster and hence improve the CPU time (i.e. making the designed tracking system to be suitable for real time implementations).
In addition, we can also reduce the errors of assigned measurements to wrong tracks (i.e. dealing with data association problem successfully).
Furthermore, both GNN and JPDA operate effectively under nominal conditions, these methods are susceptible to spoofing attacks that inject falsified detections or manipulate trajectory continuity. 

To address these unique challenges, we firstly propose to use the Graph Convolutional Neural Network for assignment learning in the GNN, and we call this tracking mechanism as trackerGNN. 
Interested readers can find more detailed designation of Graph Neural Network in \cite{le2025dpfaga, le2025HDQNN}. 
Deep learning approaches, including Graph Neural
Network and Convolutional Neural Network-based trackers, offer adaptability and learning-based generalization, especially for GNN \cite{Wang20, Wangconf2018, le2025HDQNN, Zahin19, Zahin20}.
In the following, we briefly describe the algorithm to the multiple target tracking mechanism, which is used to the JPDA. 
In fact, we employ the data correlation processing to assign the best measurements to track associations. Firstly, we will use the Kalman filter for predicting the possible future position, given the predetermined associated covariance matrix of data. 
We will evaluate the valid observation/measurements belonging to the region/cluster or not in probabilistic manner. 
The second step is to associate the measurements with the tracks. 
This is because 1) the measurement is in region of one cluster with multiple tracks or in the border of multiple clusters with multiple tracks and 2) multiple measurements are in the same region of target track. 
To address this critical issue, we develop the Hungarian matching algorithm \cite{Tan12} to determine the optimal assigned pairs of observation and target track.
\vspace{-0.1in}
\subsection{Logging and Reproducibility}
\label{Logging}
\vspace{-0.1in}
SpoofTrackBench emphasizes reproducibility through:
1) \textbf{Detection Stream Separation}: Clean and spoofed logs are stored distinctly, preserving origin and labeling integrity.
2) \textbf{Seed Management}: Each scenario initializes random seeds with logging of seed IDs per run, enabling exact reconstruction.
3) \textbf{Snapshot Evaluation}: At each timestep, tracker outputs, assignment matrices, and drift metrics are saved as MATLAB structs.
4) \textbf{Auto-Export Mechanism}: Upon scenario completion, the benchmark auto-generates comparison plots (assignment accuracy, drift overlays), summary statistics, and scenario logs under organized directories.
Visual exports include 1) assignment history overlays with spoof highlighting, 2) cluster purity and trajectory divergence plots, and 3) 2D/3D toggled views (dimMode) with tracker ID labels.
The benchmark is designed for extensibility: future tracker modules, spoof types, and multi-sensor fusion configurations can be integrated with minimal code changes using scenario template inheritance and modular helper functions.

\section{Evaluation Strategy}
\label{EvaluationStrategy}

SpoofTrackBench employs a layered evaluation framework to quantify tracker integrity, interpret spoof resilience, and compare architectural behavior under controlled adversarial conditions. The evaluation strategy spans metric logging, visual overlays, and bidirectional tracker diagnostics \cite{shabtai2016detection, zheng2021deep, rudenko2020human}.
\vspace{-0.1in}
\subsection{Performance Metrics}
\label{PerformanceMetrics}
\vspace{-0.1in}

To robustly assess tracking fidelity and spoof impact, we determine the following performance metrics.
\begin{itemize}
    \item \textbf{Drift-from-Truth Quantification}: For each tracked platform, we log mean and maximum Euclidean error between reconstructed trajectory and ground truth, per timestep.
    \item \textbf{Assignment Divergence}: Measures mismatch between assigned detections and true platform associations using normalized confusion scores and switch frequencies.
    \item \textbf{Cluster Misalignment}: Evaluates clustering purity and probability-weighted error across temporal windows, particularly sensitive to ghost spoof insertions.
    \item \textbf{Spoof Detection Flags}: Each detection is labeled, enabling downstream statistics on spoof inclusion, recovery rates, and false positive attribution.
\end{itemize}
\vspace{-0.1in}
\subsection{Cross-Tracker Comparison}
\label{CrossTracker}
\vspace{-0.1in}
SpoofTrackBench supports bidirectional comparison of tracking architectures through an automated evaluation module.
This function computes \textbf{Numerical Benchmarking}, \textbf{Visual History Comparison}, and {Spoof-Type Segmentation}.
\textbf{Numerical Benchmarking} aggregates drift errors, assignment accuracy, spoof recovery rate, and false association ratios across both tracker outputs. 
\textbf{Visual History Comparison} generates multi-timestep overlays (assignment, spoof recovery, cluster transitions) with side-by-side renderings.
Results of \textbf{Spoof-Type Segmentation} are bucketed by spoof types (i.e. drift, ghost and mirror), enabling conditional analysis and architecture-specific vulnerability scoring.
Exported examples include 1) drift error heatmaps across platforms and timesteps, 2) cluster purity timelines and divergence plots, and 3) assignment history and switch rate graphs.
\section{Experiments, Results and Discussion}
\label{ExperimentsResults}
We present simulation results designed to evaluate the spoof resilience, assignment fidelity, and interpretability of tracking architectures using SpoofTrackBench. Experiments span varied spoof types, tracker configurations, and visualization diagnostics.
\subsection{Simulation Setup}
\label{SimSetup}
We simulate the scenario of drones flying and get the data for the radar measurements.
This dataset would be used for our experiment on target tracking in the subsequent sections.

\noindent \textbf{Scenario Setup}: 
For the initialization of the scenario, simulation time is run until objects halt movement. 
Three moving and one stationary objects are initialized each with a class ID. 
In addition to the dimensions of moving object being the same, the radar cross section signature (RCS),  whose parameters decide the intensity of the reflected radar signal from the target of each moving object are the same as well. 
Used parameters for the RCS signature include signature fluctuation model, the sampled RCS pattern, pattern frequency, Azimuth angles and elevation angles. 
The trajectories of each moving object vary in course, ground speed, climb rate, auto-pitch and auto-bank for a diverse readings.  
Experiments were orchestrated using a cyclic spoof evaluation loop, where each run initializes a unique spoof configuration:
1) \textbf{Spoof Type Cycling}: Drift, ghost, and mirror spoof categories are applied across multiple scenario seeds using the benchmark’s orchestration module;
2) \textbf{Parameter Variation}: Detection noise levels, spoof injection timing, and agent counts were varied across runs to assess robustness under dynamic stress conditions;
3) \textbf{Platform Configuration}: Both cooperative and non-cooperative platforms were simulated. Cooperative agents use ADSB overlays, while radar-only detections were reserved for spoof-prone entities;
4) \textbf{Runtime Logging}: Each evaluation run generates a structured folder tree containing tracker outputs, spoof injection parameters, and interpretability exports.
Fusion center inputs are normalized across sensors, maintaining dimensional consistency and detection field integrity across both clean and spoofed streams.
\vspace{-0.1in}
\subsection{GNN vs. JPDA without Spoofing Attacks}
\begin{figure*}[ht]
\centering
\vspace{-0.3in}
\mbox{\subfigure[]{\includegraphics[width=2.0in]{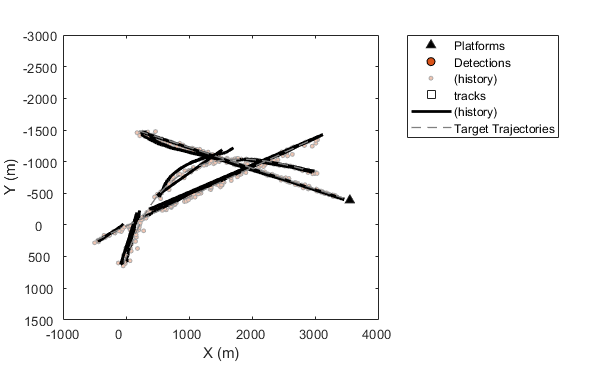}\label{Fig_TrajGNN} }
\subfigure[]{\includegraphics[width=2.0in]{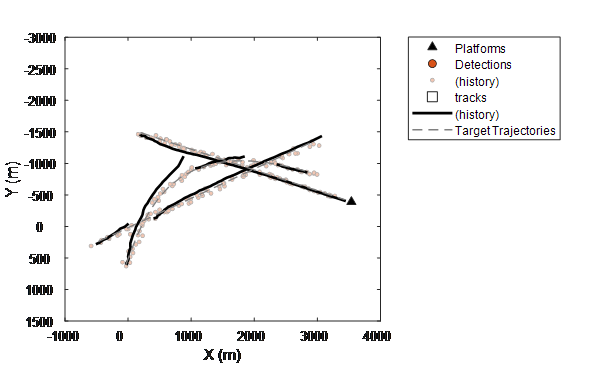} 
\label{Fig_Traj}}
}
\vspace{-0.2in}
\caption{Trajectory tracking methods: (a) The GNN trajectory tracking, and (b) The JPDA trajectory tracking.}
\label{Fig4DQNN_RFEntanglement}
\vspace{-.3in}
\end{figure*}

\begin{figure*}[ht]
\centering
\vspace{-0.3in}
\mbox{\subfigure[]{\includegraphics[width=2.0in]{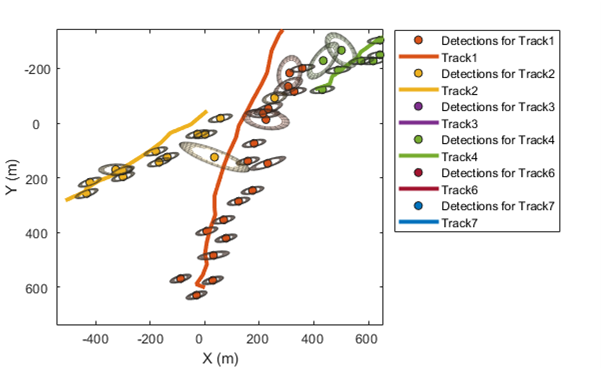}\label{Fig_Traj2} }
\subfigure[]{\includegraphics[width=2.0in]{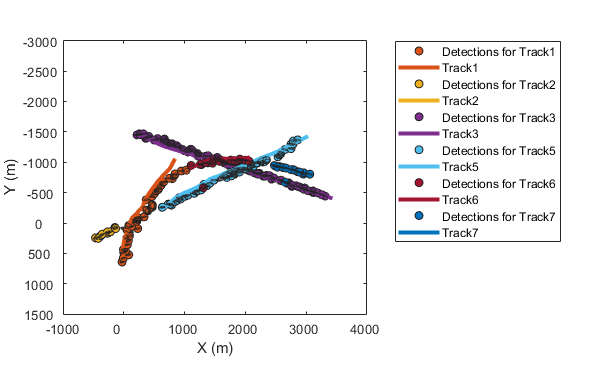} 
\label{Fig_Traj2JPDA}}
}
\vspace{-0.2in}
\caption{Evaluation of trackers: (a) Evaluation of GNN trajectory tracker, and (b) Evaluation of JPDA trajectory tracker.}
\label{Fig4DQNN_RFEntanglement1}
\vspace{-.3in}
\end{figure*}

\vspace{-0.1in}
In the following, we present some essential results.
Firstly, we consider the scenario of trajectories, where each moving object vary in course, ground speed, climb rate, auto-pitch and auto-bank for a diverse readings.
The mentioned scenario can be seen in Fig. \ref{Fig_TrajGNN}, which displays the detections (history) along with detection trajectories in the solid black lines and true target trajectories in dotted grey lines.
In Figs. \ref{Fig_Traj2} and \ref{Fig_Traj2JPDA}, we evaluate the GNN-based and JPDA-based multi-object tracking algorithms, where detections in colored dots along with their probability distribution surrounding them are displayed with predicted trajectories in solid colored lines. 
It can be seen that when objects are sparsely spaced, i.e. track one and two from -600 to 0 meters along the x-axis, trajectory predictions are well defined. From 0 to 600 along the x-axis, however, tracked objects are cluttered and as a result trajectories and assignments becomes disorganized. 
While Fig. \ref{Fig_TrajGNN} maps trajectories similar to Fig. \ref{Fig_Traj}, it can be seen that it fails at overlapped tracks.
In a multi-object tracking algorithm, detections must assigned to a tracked object without the interference of other objects detection, meaning the prediction of one object should not be affected by the observations of another. 
GNN provides an ease of computation, however, its performance is limited with cluttered objects. 
JPDA can overcome the GNN's cluster issues by assigning weights to the closest detection of each tracked object and using the weight combinations for assignment. 
Therefore, when a cluster of objects is established, each detection is not discretely assigned to each object, but is collaboratively performed with the adaptive weight.
Hence, JPDA has a higher accuracy of assignment when comparing GNN's assignment to the graph of the original object trajectories. 
\vspace{-0.1in}
\subsection{GNN vs. JPDA under Spoof Pressure}
\label{TrackerComparison}
\vspace{-0.1in}
\begin{figure*}[ht]
\centering
\vspace{-0.3in}
\mbox{\subfigure[]{\includegraphics[width=2.0in]{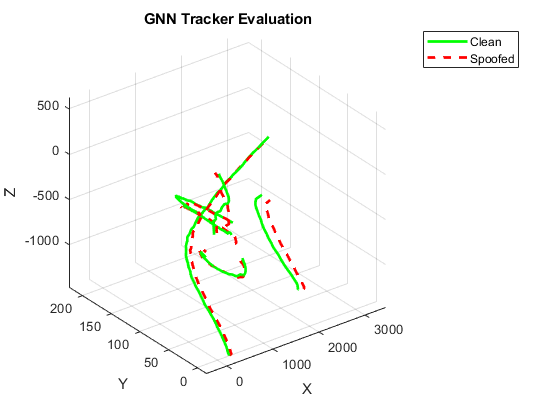}\label{Fig_Traj3_SpoofGNN} }
\subfigure[]{\includegraphics[width=2.0in]{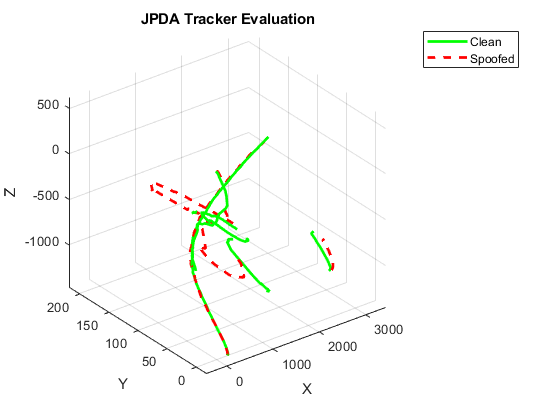} 
\label{Fig_Traj3_SpoofJPDA}}
}
\vspace{-0.2in}
\caption{Evaluation of trackers under Ghost Spoofing: (a) Evaluation of GNN trajectory tracker, and (b) Evaluation of JPDA trajectory tracker.}
\label{Fig4DQNN_RFEntanglement2}
\vspace{-.3in}
\end{figure*}
We firstly evaluate tracking performances of the GNN and JPDA under the consideration of spoofing attacks in Figs. \ref{Fig_Traj3_SpoofGNN} and \ref{Fig_Traj3_SpoofJPDA}, respectively.
In Ghost Spoofing Attacks, the spoofing attacks cause significant distortions in the formulated trajectories of both GNN and JPDA trackers, which manifest as degraded track continuity, increased positional drift, and reduced assignment confidence. 
These effects are especially pronounced in JPDA due to its probabilistic association mechanism, which is more susceptible to spoof-induced ambiguity within the gating region.
GNN exhibits stronger performance in spatial reasoning, especially under ghost and mirror spoof types. Its message-passing architecture supports cross-detection affinity scoring, enabling partial immunity to clustering confusion and ghost proliferation.
GNN uses hard gating and assigns each detection to the nearest track if it falls within a predefined threshold. Spoofed detections outside the gate are ignored. 
Even if spoofed detections are inside the gate, only the closest one is selected, making GNN more resilient to clutter and spoof bursts.
However, GNN trackers exhibit:
1) Increased sensitivity to irregular temporal cadence (e.g. intermittent spoof injection),
2) Reduced explainability due to latent node embeddings and opaque decision boundaries.
\begin{figure*}[ht!]
\centering
\vspace{-0.3in}
\mbox{
\subfigure[Drift Spoof]{\includegraphics[width=1.2in]{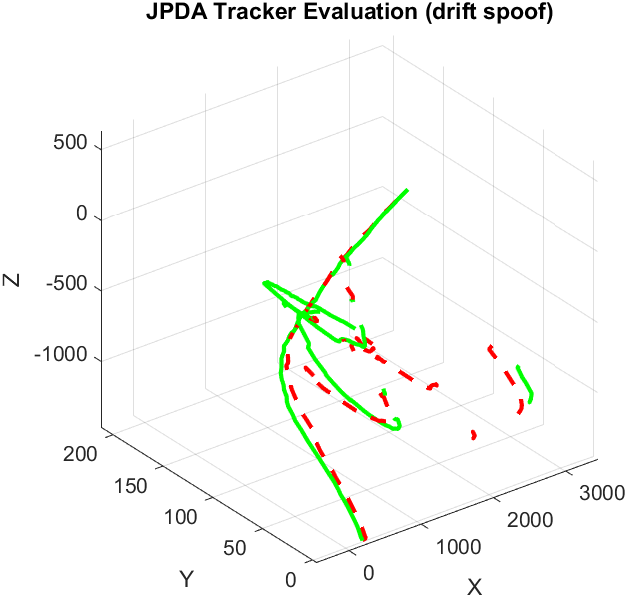}\label{TrackComparison_JPDA_drift_clean}}
\hspace{0.1in}
\subfigure[Ghost Spoof]{\includegraphics[width=1.2in]{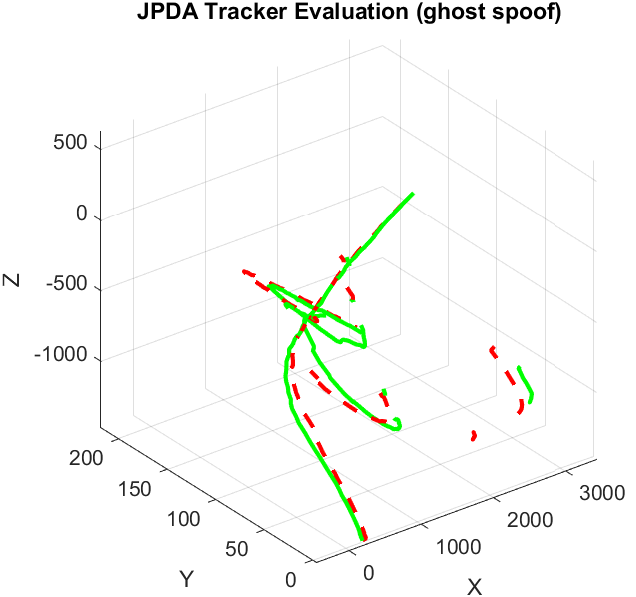}\label{TrackComparison_JPDA_ghost_clean}}
\hspace{0.1in}
\subfigure[Mirror Spoof]{\includegraphics[width=1.2in]{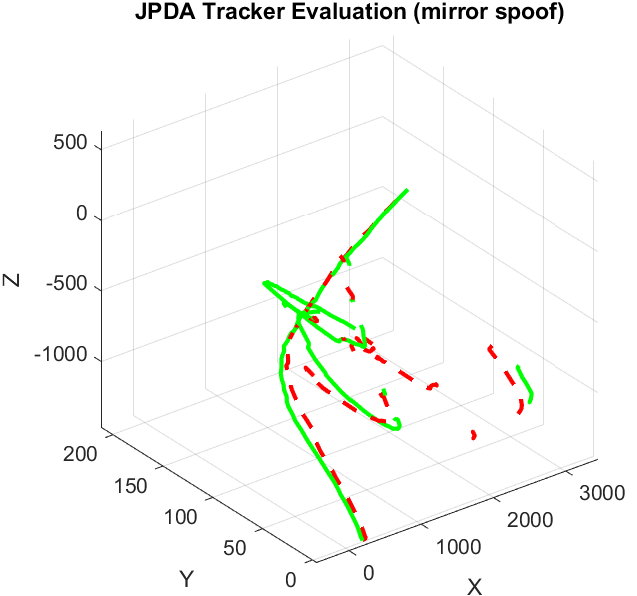}\label{TrackComparison_JPDA_mirror_clean}}
}
\vspace{-0.1in}
\caption{JPDA tracker performance under spoof injection scenarios (solid green lines represent estimated UAV trajectories, while dashed red lines illustrate spoofed paths affected by gradual drift injection). 
(a) Drift spoof, 
(b) Ghost spoof,
(c) Mirror spoof. 
}
\label{Fig4DQNN_RFEntanglement3}
\vspace{-0.3in}
\end{figure*}
\begin{table*}[!ht]
\centering
\caption{Spoof Impact Benchmark: Drift (m) and Normalized Impact (\%)}
\begin{tabular}{|l| l| c| c|}
\toprule
\textbf{Tracker} & \textbf{Spoof Type} & \textbf{Drift (m)} & \textbf{Normalized Impact (\%)} \\
\midrule
GNN   & drift  & 77.10 & 15.42 \\
GNN   & ghost  & 66.25 & 13.25 \\
GNN   & mirror & 72.10 & 14.42 \\
JPDA  & drift  & 66.10 & 13.22 \\
JPDA  & ghost  & 77.05 & 15.41 \\
JPDA  & mirror & 75.38 & 15.08 \\
\midrule
\textbf{Average (GNN)}           & -- & 72.48 & 14.36 \\
\textbf{Average (JPDA)}          & -- & 72.84 & 14.90 \\
\textbf{Average (drift spoof)}   & -- & 71.60 & 14.32 \\
\textbf{Average (ghost spoof)}   & -- & 71.65 & 14.33 \\
\textbf{Average (mirror spoof)}  & -- & 73.74 & 14.75 \\
\textbf{Overall Average}         & -- & 76.24 & 15.25 \\
\bottomrule
\end{tabular}
\vspace{-0.3in}
\label{tab:spoof_impact}
\end{table*}
In contrast, JPDA uses soft gating and computes association probabilities $\beta_{ij}$ for each detection-track pair. When spoofed detections fall within the gate, they dilute the probability mass, reducing confidence in true associations. This can lead to track drift, fragmentation, or premature termination.
In JPDA, the association probability $\beta_{ij}$ is computed as $\beta_{ij} = \frac{P(\mathbf{z}_t^{(i)} \mid \text{track } j)}{\sum_{k=1}^{M} P(\mathbf{z}_t^{(k)} \mid \text{track } j)}$, where $M$ is the number of detections in the gate.
When spoofed detections are injected, the denominator increases, reducing $\beta_{ij}$ for the true detection.
So, the tracker may update the track with a weighted average that includes spoofed positions.
Over time, this leads to confidence decay, trajectory corruption, and false track coalescence.
Therefore, JPDA offers the following benefits depending on the considered environment.
1) Transparent assignment logic with interpretable gating and likelihood updates.
2) Greater resilience under gradual drift scenarios through Kalman filter continuity.
3) Weaker discrimination under dense spoof scenarios due to probabilistic dilution and clutter overlap.
Figs.~\ref{TrackComparison_JPDA_drift_clean}, \ref{TrackComparison_JPDA_ghost_clean}, and \ref{TrackComparison_JPDA_mirror_clean} provide detailed visualizations of JPDA tracker performance under three spoofing scenarios. 
In Fig.~\ref{TrackComparison_JPDA_drift_clean}, solid green lines represent the estimated UAV trajectories, while dashed red lines illustrate spoofed paths affected by gradual drift injection, highlighting JPDA’s ability to preserve track continuity despite adversarial corruption. 
Fig.~\ref{TrackComparison_JPDA_ghost_clean} shows tracked UAV paths in green and false detections introduced by ghost spoofing in red, demonstrating JPDA’s resilience in filtering out non-physical targets. 
In Fig.~\ref{TrackComparison_JPDA_mirror_clean}, green lines depict estimated UAV trajectories, while red dashed lines reflect spoofed paths generated by symmetry-based mirror injection. 
Collectively, these visualizations underscore JPDA’s robustness in maintaining trajectory integrity, rejecting deceptive inputs, and resisting spoof-induced distortions across diverse attack types.
Finally, these findings suggest complementary deployment strategies, i.e. JPDA for low-spoof environments with clearer temporal continuity, and GNN for adversarial scenarios requiring dynamic assignment discrimination.
\vspace{-0.1in}
\subsection{Spoof-Type Impact on Cluster Integrity}
\label{SpoofImpact}
\vspace{-0.1in}
Table \ref{tab:spoof_impact} demonstrates the performances of GNN and JPDA trackers across all spoof types, each exhibiting distinct clustering behaviors. 
1) \textbf{Drift Spoofs}: Induce gradual trajectory bending, resulting in delayed cluster separation and subtle assignment drift. 
2) \textbf{Ghost Spoofs}: Generate phantom detections that infiltrate clusters, triggering early confusion and tracker instability.
3) \textbf{Mirror Spoofs}: Create symmetrical echoes that challenge spatial models, often producing mirrored clusters with high affinity misgroupings.
These observations highlight the unique failure modes introduced by different spoofing strategies and underscore the importance of spoof-aware evaluation in resilient tracking pipelines.
GNN’s affinity-based graph edges partially resolve ghost confusion but can misassign in drift and mirror cases. JPDA maintains cluster purity longer during drift but is prone to ghost contamination. This underscores the need for spoof-specific tracker evaluation rather than generic benchmarking.
\vspace{-0.1in}
\subsection{Spoof Detection Flags: UAV Trajectory Overlay under Drift Injection}
\vspace{-0.1in}
\begin{figure}[ht!]
\vspace{-0.3in}
    \centering
    \includegraphics[width=0.35\textwidth]{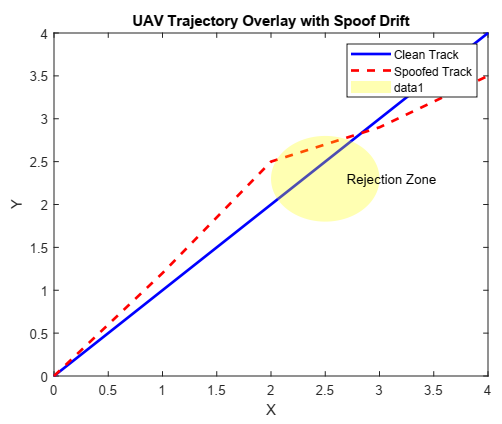}
    \vspace{-0.1in}
\caption{UAV Trajectory Overlay with Spoof Drift. The clean track (solid blue) represents the true UAV trajectory, while the spoofed track (dashed red) illustrates the adversarial drift injection. The yellow rejection zone highlights regions where spoofed detections are filtered by the tracking system.}    \label{Detection}
\vspace{-0.3in}
    \end{figure}
Fig.~\ref{Detection} illustrates the UAV trajectory overlay under spoof drift injection, where the clean track (solid blue) and spoofed track (dashed red) diverge across the X-Y plane. The yellow rejection zone delineates regions where spoofed detections are actively filtered by our proposed detection scheme. This mechanism enhances tracking resilience by isolating corrupted inputs and preserving trajectory integrity.
By leveraging spatial divergence and rejection logic, our spoof-aware detection framework mitigates the risk of tracking failure under adversarial drift attacks. It enables robust trajectory recovery, improves downstream tracker performance, and supports interpretable diagnostics for spoof injection events. This capability is critical for mission-sensitive UAV operations in contested environments.
\section{Conclusion and Future Work}
\label{ConclusionFutureWork}
SpoofTrackBench establishes a reproducible, architecture-agnostic benchmark for radar-based tracking under adversarial spoof conditions. By integrating modular spoof injection, tracker comparison, and interpretability overlays, it provides the community with an open framework for evaluating assignment resilience and trajectory fidelity.
Key contributions include 1) a flexible spoof injection pipeline supporting drift, ghost, and mirror types with configurable parameters; 2) architecture-neutral benchmarking modules for classical JPDA and GNN trackers, enabling quantitative and qualitative comparison under identical spoof scenarios; and 3) automated logging, evaluation snapshots, and interpretability overlays to support scientific reproducibility and visual storytelling.

To further advance spoof resilience and tracking interpretability, we envision the following roadmap:
1) \textbf{Integration of Deep Neural Spoof Classification (SpoofNet)}:  
    We will embed SpoofNet with recent AI/ML algorithms \cite{le2025dpfaga, le2025HDQNN, Wang20, Tan2021, Tan19a, Zahin19, Zahin20, Tan18b}, a real-time spoof detection module trained on injection-labeled examples (drift, ghost and mirror), to classify detection integrity on-the-fly. This module will enhance interpretability overlays with spoof confidence scores and support dynamic tracker behavior based on spoof severity.
2) \textbf{Hybrid Deep Quantum Neural Network (DQNN) Tracker Architecture}:  
    Inspired by JPDA's probabilistic foundations and GNN’s spatial modeling, we propose a DQNN tracker integrating SpoofNet’s outputs with quantum gate-inspired assignment logic \cite{le2025HDQNN}. This architecture will probabilistically reason about detection ambiguity while maintaining stable tracking in high-spoof environments.
    
\vspace{10pt}
\noindent\textbf{Acknowledgment}: 
This work was supported in part by the Commonwealth Cyber Initiative (CCI) Experiential Learning Program and the South Big Data Innovation Hub Partnership Nucleation program.

\vspace{-10pt}
\bibliographystyle{IEEEtran}
%
\bibliography{references}

\end{document}